\documentclass[prl,twocolumn,superscriptaddress,nobibnotes]{revtex4-2}
\usepackage{graphicx}
\usepackage{amssymb}
\usepackage{amsmath}
\usepackage{bm}
\usepackage{color}
\usepackage{float}
\usepackage{setspace}
\usepackage{physics}
\usepackage{subfigure}
\usepackage{soul}
\usepackage{lineno}

\begin{document}
%\linenumbers
\title{Switching off microcavity polariton condensate near the exceptional point}

\author{Yao Li}
\affiliation{Tianjin Key Laboratory of Molecular Optoelectronic Science, Institute of Molecular Plus, School of Science, Tianjin University, Tianjin 300072, China} 

\author{Xuekai Ma}
\email{xuekai.ma@gmail.com}
\affiliation{Department of Physics and Center for Optoelectronics and Photonics Paderborn (CeOPP), Universit\"{a}t Paderborn, Warburger Strasse 100, 33098 Paderborn, Germany}

\author{Zaharias Hatzopoulos}
\affiliation{Institute of Electronic Structure and Laser (IESL), Foundation for Research and Technology-Hellas (FORTH), Heraklion 71110, Greece}

\author{Pavlos Savvidis}
\affiliation{School of Science,Westlake University, 18 Shilongshan Road, Hangzhou 310024, Zhejiang Province, China}
\affiliation{Institute of Natural Sciences, Westlake Institute for Advanced Study, 18 Shilongshan Road, Hangzhou 310024, Zhejiang Province, China}
\affiliation{Institute of Electronic Structure and Laser (IESL), Foundation for Research and Technology-Hellas (FORTH), Heraklion 71110, Greece}
\affiliation{Department of Nanophotonics and Metamaterials, ITMO University, St. Petersburg 197101, Russia
}

\author{Stefan Schumacher}
\affiliation{Department of Physics and Center for Optoelectronics and Photonics Paderborn (CeOPP), Universit\"{a}t Paderborn, Warburger Strasse 100, 33098 Paderborn, Germany}
\affiliation{Wyant College of Optical Sciences, University of Arizona, Tucson, AZ 85721, USA}

\author{Tingge Gao}
\email{tinggegao@tju.edu.cn}
\affiliation{Tianjin Key Laboratory of Molecular Optoelectronic Science, Institute of Molecular Plus, School of Science, Tianjin University, Tianjin 300072, China} 

\begin{abstract}
{\textbf{Gain and loss modulation are ubiquitous in nature. An exceptional point arises when both the eigenvectors and eigenvalues coalesce, which in a physical system can be achieved by engineering the gain and loss coefficients, leading to a wide variety of counter-intuitive phenomena. In this work we demonstrate the existence of an exceptional point in an exciton polariton condensate in a double-well potential. Remarkably, near the exceptional point, the polariton condensate localized in one potential well can be switched off by an additional optical excitation in the other well with very low (far below threshold) laser power which surprisingly induces additional loss into the system. Increasing the power of the additional laser leads to a situation in which gain dominates in both wells again, such that the polaritons re-condense with almost the same density in the two potential wells. Our results offer a simple way to optically manipulate the polariton condensation process in a double-well potential structure. Extending such configuration to complex potential well lattices offers exciting prospects to explore high-order exceptional points and non-Hermitian topological photonics in a non-equilibrium many-body system.}}
\end{abstract}
\maketitle

%\section{Introduction}

Effective Hamiltonians of non-Hermitian nature play a crucial role in our understanding of a plethora of modern physical systems \cite{1Nimrod-non hermitian book, 2Konotop Rev}. This is in spite of the underlying common principle that in quantum mechanical systems Hermiticity is required to keep a real-valued spectrum of eigenvalues. Complex eigenvalues generally appear in systems governed by non-Hermitian Hamiltonians. However, Bender et.al. \cite{3Bender 1998prl} observed that under one special condition systems can still sustain a real-valued energy spectrum. That is if the Hamiltonian commutes with the \textit{PT} operator, where \textit{P} and \textit{T} are the parity and time-reversal operator, respectively. In recent years photonic platforms are widely employed to study \textit{PT} symmetry in view of the fact that the refractive index distribution can be judiciously engineered, that is, the real part of the refractive index can be tailored to be an even function whereas the imaginary part remains an odd function such that \textit{PT} symmetry can be fulfilled. A striking phase transition can occur at a point in parameter space (the symmetry breaking point or exceptional point) where the real eigenvalues become complex-valued. This transition can be induced by modulation of the real or imaginary components of the external potential in the effective Hamiltonian. Near this exceptional point, a large variety of counter-intuitive phenomena has been reported, for example, \textit{PT} symmetric synthetic lattices \cite{4PT synthetic lattice} or microcavities \cite{5PT yanglan}, single mode \cite{6single mode laser xiangzhang, 7single mode laser CDN} and vortex lasing \cite{8PT vortex laser} in a ring cavity, unidirectional light propagation in a waveguide \cite{9unidirectional PT}, suppression and revival of lasing due to loss in coupled whispering-gallery-mode microcavities \cite{10loss revival  yanglan}, improved sensitivity in single or coupled resonators \cite{11high order EP CDN, 12high order EP yanglan}, optical isolation in \textit{PT} symmetric microcavities \cite{13PT-xiaomin}, formation of an exceptional ring in a photonic crystal \cite{14exceptional circle}, and an enhanced linewidth of a phonon laser mode at the exceptional point \cite{15phonon laser at EP}. In contrast to regular laser operation, a non-Hermitian system can behave surprisingly near the exceptional point where changes in gain can shut down or re-excite lasing as demonstrated in coupled resonators \cite{16reverse pump experiment, 17reverse pump theory, 18lasing termination}. Approaching the exceptional points in these works is achieved using asymmetrical pumping schemes. 

\begin{figure}[t]
 \centering
 \includegraphics[width=\linewidth]{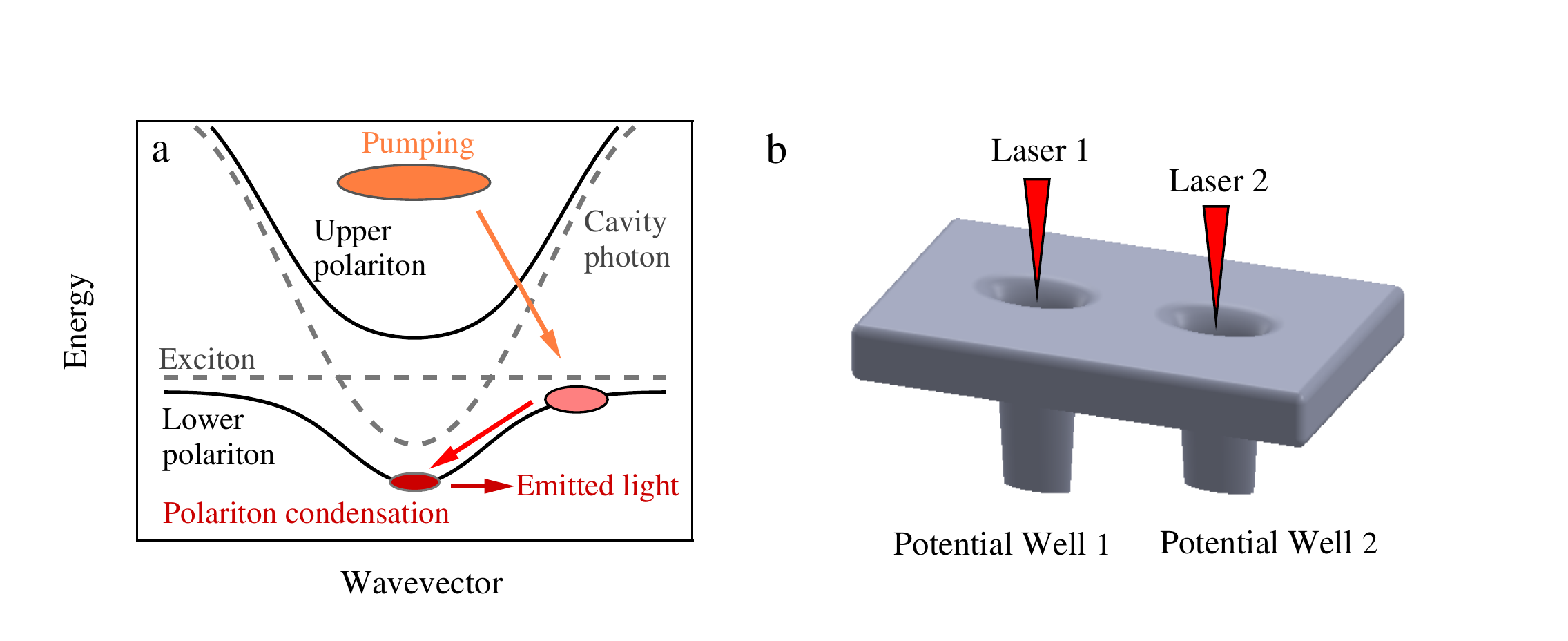} 
 \caption{\textbf{Excitation configuration.} Schematics of polariton condensation process in the planar microcavity (a) and the structure of a built-in double-well potential (b) which is pumped by two independent non-resonant laser beams.} \label{fig:dispersion}
\end{figure}

Exciton polaritons form as hybrid particles from excitons in a quantum well that strongly couple with the photon mode of a planar optical resonator. In an excitation regime where polaritons behave like bosonic quasi-particles, they can experience condensation and show spontaneous macroscopic coherence under non-resonant excitation \cite{20Deng Science, 21Nature CdTe, 22Balili Science}, as shown in Fig. \ref{fig:dispersion}(a). Thanks to the spontaneous decay of photons from the microcavity, polariton systems are inherently driven-dissipative in nature and as such  provide an excellent platform to study \textit{PT} symmetry \cite{23Ma NJP} and non-Hermitian physics. In our previous works \cite{24Gao Nature, 25Gao PRL} we demonstrated that the polariton condensate wavefunction and energy distribution can be strongly modified by the gain and loss coefficients in an optical billiard. Also in an exciton polariton system in coupled microresonators the influence of pump asymmetry was explored and it was demonstrated that relaxation kinetics plays a crucial role in the condensation process \cite{19polariton dimer}. Polariton dimers also offer a platform to study polariton-polariton nonlinear interactions with Josephson oscillations, quantum self trapping \cite{26J Bloch Josephson and trapping}, interplay of interference and nonlinearity \cite{27hopping phase}, and bistability \cite{28J Bloch coupled bistable}. However, the existence of an exceptional point and its role for the polariton condensation was not demonstrated or discussed in such structures.

In the present work we utilize a particularly simple and robust double-well potential structure with slightly asymmetric sites as sketched in [Fig. \ref{fig:dispersion}(b)](the details of the sample can be seen in the experimental section). We show that for non-resonant optical excitation polariton condensates can be loaded into the double-well in a controlled manner , and demonstrate that an exceptional point can be realized and explored in detail varying the optical excitation parameters. In our experiment, we use two independent non-resonant pumping lasers to separately excite the two potential wells and find that the polariton condensation in one potential well can be altered dramatically and even shut down completely by only controlling the pump intensity on the adjacent potential well (in this case the double-well potential system as a whole thus falls below the condensation threshold). The reason is that varying the pump intensity on one of the potential wells changes the coupling of them, giving rise to the appearance of an exceptional point which results in substantial redistribution of the modes and reduced effective gain for polaritons in the coupled wells. With further increasing the pump intensity on the adjacent well, the polariton dimer starts condensation again as the system is tuned away from the exceptional point and bifurcates into two modes, antibonding and bonding. Our results show that the exceptional point can be easily tailored in such kind of macroscopic quantum system and our work can be systematically extended to 1D or 2D lattices to investigate anomalous edge modes based on polariton condensates \cite{Lee edge non hermitian, topological invariants  wangzhong}. 

\section{r\lowercase{esults}}
\section*{T\lowercase{heory of non-Hermitian degeneracy}}
In theory, the Hamiltonian of a two-level non-Hermitian system like a coupled polariton dimer trapped in two potential wells 1 and 2 can be expressed as follows:

\begin{figure}[t]
 \centering
 \includegraphics[width=\linewidth]{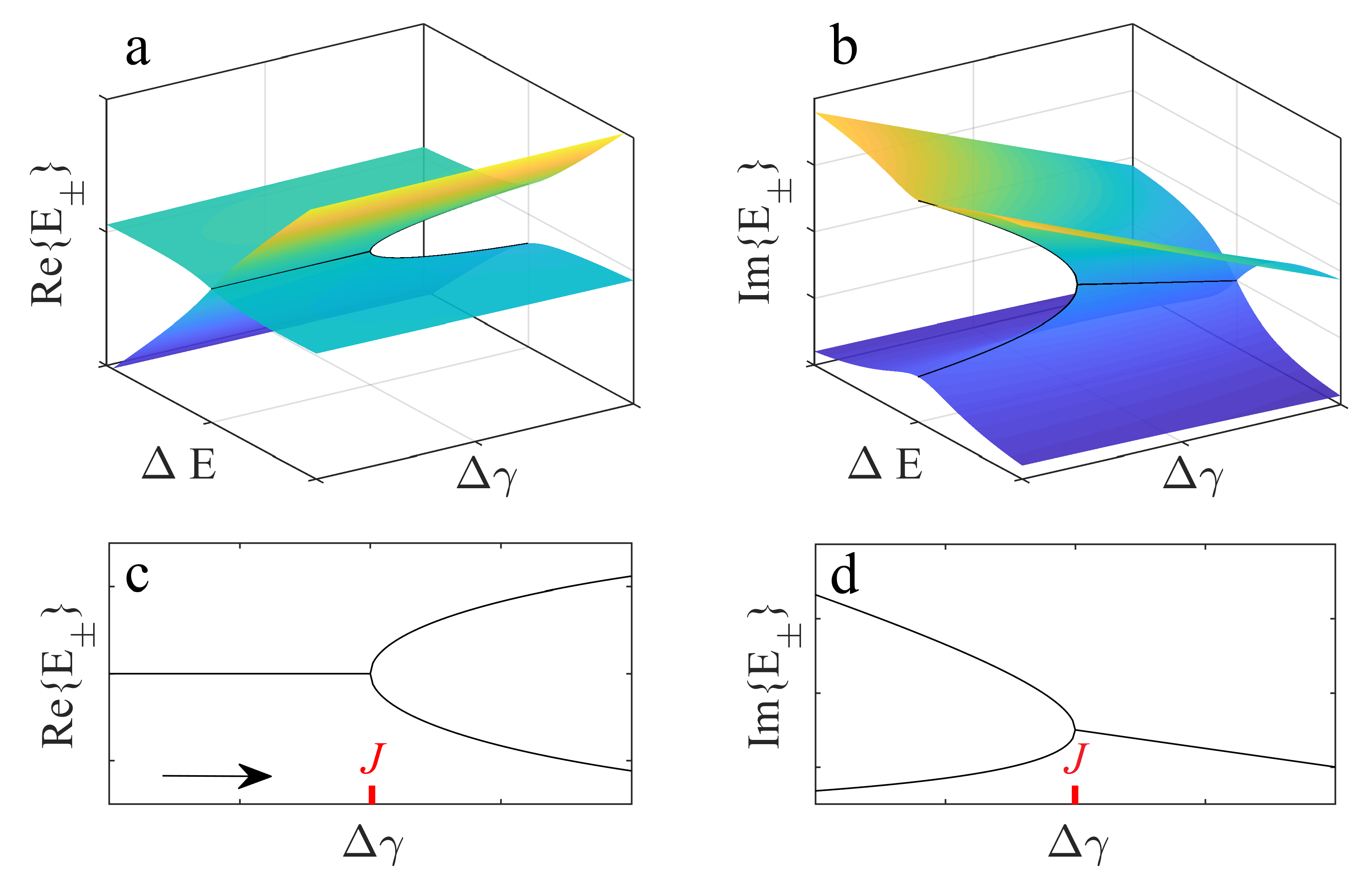} 
 \caption{\textbf{Eigenenergies of a two-level non-Hermitian Hamiltonian near an exceptional point.} (a) Real and (b) imaginary parts of the eigenenergy in a two dimensional parameter space, where $\Delta\gamma$ = $\varGamma_1-\varGamma_2$, $\Delta E$= $E_1-E_2$. Relations between the (c) real and (d) imaginary parts of the eigenenergy with $\Delta\gamma$ along the route $\Delta{E}=0$ highlighted in (a) and (b). The arrow indicates the decrease of $\Delta{\gamma}$. The exceptional point appears at $\Delta{\gamma}=J$. } \label{fig:eigenenergy}
\end{figure}

\begin{equation}
H
=
\left[
\begin{array}{ccc}
    E_1+i \varGamma_1 & J/2 \\
    J^*/2 & E_2+i \varGamma_2 \\
\end{array}
\right]
\end{equation}
where $E_1$($\varGamma_1$) and $E_2$($\varGamma_2$) correspond to the real (imaginary) part of the polariton energies in the two potential wells, J is the coupling strength between the two wells. The eigenenergies of the Hamiltonian are $E_\pm=(E_1+E_2+i\varGamma_1+i\varGamma_2)/2\pm\sqrt{(J^2+[E_1-E_2+i\varGamma_1-i\varGamma_2]^2)}/2$. The dependence of the real and imaginary parts of the eigenenergies on $\Delta{E}=E_1-E_2$ and $\Delta{\gamma}=\Gamma_1-\Gamma_2$ are shown in Figs. \ref{fig:eigenenergy}(a) and \ref{fig:eigenenergy}(b). 
%To simplify the discussion, we assume that $E_1$=$E_2$ and $\varGamma_1$\neq$\varGamma_1$. 
We introduce asymmetric pumping of the polariton dimer by sequentially exciting the two potential wells: the potential well 1 is excited above the threshold, then the pumping power of the potential well 2 is increased from zero. Before the system approaches the exceptional point with the difference of the gain level of the two potential wells $\Delta\gamma>J$ (the potential well 1 has a larger gain than 2) along the route at $\Delta{E}=0$ shown in Figs. \ref{fig:eigenenergy}(a) and \ref{fig:eigenenergy}(b), one of the two eigenstates of the above non-Hermitian Hamiltonian lies above the condensation threshold with large gain [see the lower branch in Fig. \ref{fig:eigenenergy}(d)] and another is below the threshold with large loss rate [see the upper branch in Fig. \ref{fig:eigenenergy}(d)] \cite{16reverse pump experiment}. When the pump power on the other potential well increases, the difference of the gain level diminishes as shown in Fig. \ref{fig:eigenenergy}(b) and (d) until $\Delta\gamma=J$, thus approaches the exceptional point \cite{18lasing termination}, where $J^2+[E_1-E_2+i\varGamma_1-i\varGamma_2]^2=0$. In this case, the polaritons feel a larger effective loss when the gain/loss ratio in the coupled wells approaches a critical value, which can leads to the already condensed polariton being switched off. As a consequence, the polariton dimer is below the threshold and the emission intensity is reduced greatly. When $\Delta{\gamma}$ decreases further, the system is pulled away from the exceptional point and the gain/loss contrast is reduced, thus the real parts of the eigenvalues of the system are repelled away from each other [see the bifurcation in Fig. \ref{fig:eigenenergy}(c)] and finally the polariton dimer can condensate again. The switching off of the polariton dimer can also occur when ${E_1}\neq{E_2}$ where the anticrossing of the eigenenergy of the system will appear \cite{17reverse pump theory, 16reverse pump experiment, 18lasing termination}.

\section*{E\lowercase{xperimental realization}}
In the experiment, we use a microcavity which contains 12 GaAs quantum wells in the cavity with the Rabi splitting of around 9.2 meV and is cooled down at around 4 K. The two potential wells with the distance of around 5 $\mu$m [see the sketch in Fig. \ref{fig:dispersion} (b)] are formed unintentionally during the growth process due to some defects which induce loss into the system. Such kind of potential wells can also be tailored with microstructuring mesas into the planar microcavity. From the measurement taken under low pumping power at the potential wells and planar microcavity (shown in the Supplemental Materials), we find the potential depth of the two potential wells are 1.46 meV and 1.30 meV, respectively, as demonstrated in Fig. S1 (b) and (c). This double-well is chosen which allows for realizing equal energy distribution in the two potential sites with deep well above threshold and shallow well below threshold, and polaritons in the system experience large gain level difference. All the experiments are taken using a CW laser (wavelength: 750 nm) and the laser beam is chopped with a mechanical chopper with the duty circle of 5\% to reduce heating.

\begin{figure}[t]
 \centering
 \includegraphics[width=\linewidth]{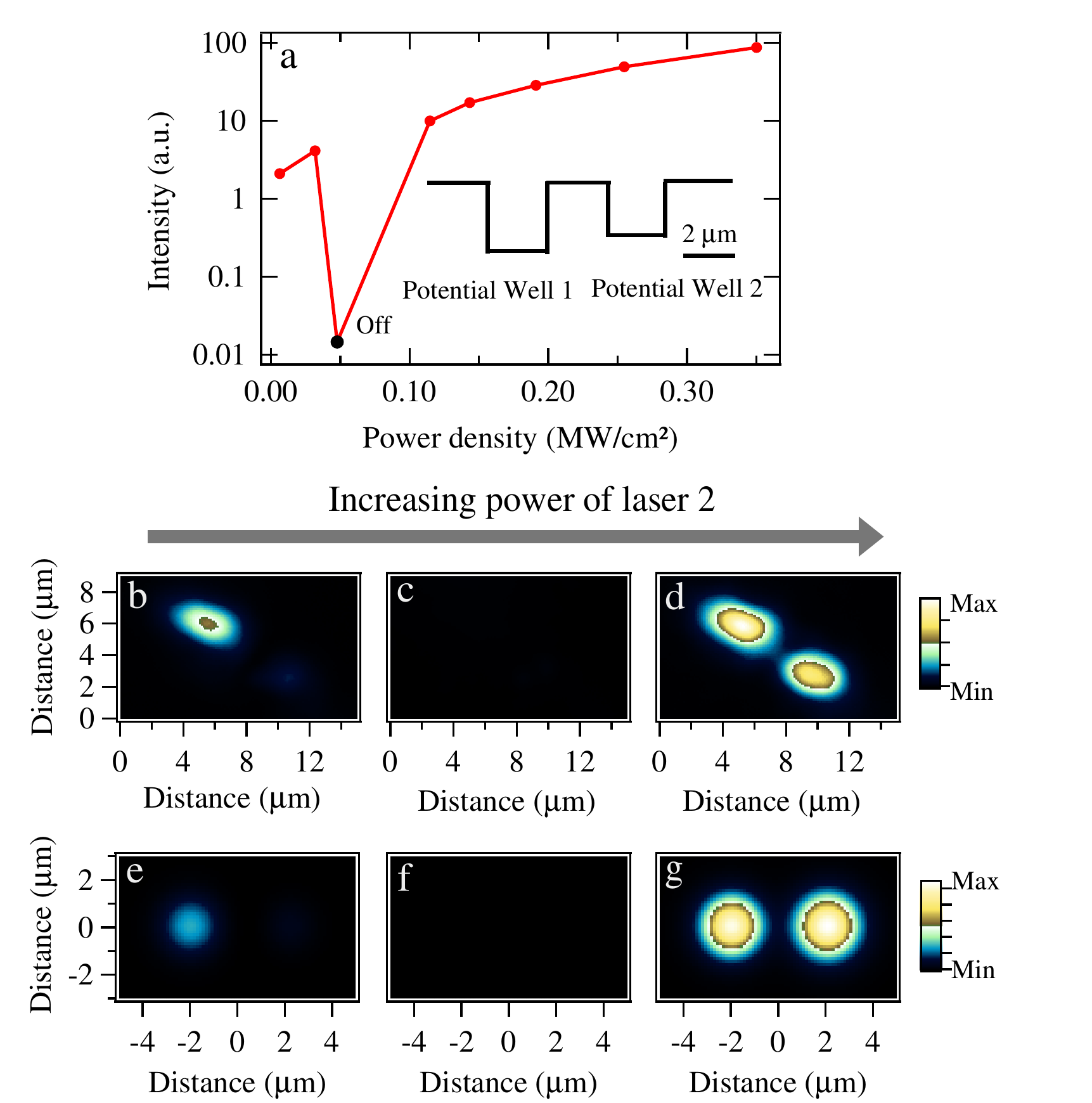} 
 \caption{\textbf{Switching off exciton polariton condensates.} (a) The emission intensity of the polariton molecule as the function of the pumping intensity of the potential well 2 with the potential well 1 being continuously excited. The real space distribution of polaritons is taken at the pump densities of 0.006MW/cm$^{2}$ (b), 0.048MW/cm$^{2}$ (c), and 0.115MW/cm$^{2}$ (d) of the second laser beam. (e-g) Time-integrated density profiles of the polariton condensates from numerical simulations at $P_1=$12 $\mu$m$^{-2}$~ps$^{-1}$ and different $P_2$: (e) $P_2=$ 0 $\mu$m$^{-2}$~ps$^{-1}$, (f) $P_2=$ 
5 $\mu$m$^{-2}$~ps$^{-1}$, and (g) $P_2=$ 15 $\mu$m$^{-2}$~ps$^{-1}$. Note that the density profiles in the experiment (b-d) and in the simulation (e-g) have different orientations.} \label{fig:density}
\end{figure}

%[as a reference, the dispersion of the planar microcavity, outside the potential wells, can be seen in Fig. S1 (a)]. 
%The detuning of the double-well is around -7 meV, estimated from the fitting of the dispersion below the threshold.]

Firstly we focus the first laser beam onto the potential well 1 with the spot size of around 2 $\mu$m.  At the pump density of around 0.45 MW/cm$^2$ ($\sim1.2P_{th}^1$, where $P_{th}^1$ is the condensation threshold of the potential well 1), the polaritons condense and are mainly located in the driven potential well 1 [see Fig. \ref{fig:density} (b)] with the energy of 1.5056 eV (the data showing the polariton condensation can be seen in Fig. S2 of the SM). Under quasi-resonant pumping, similar localization of exciton polaritons in a photonic dimer due to quantum self localization was observed in \cite{26J Bloch Josephson and trapping}, and polaritons can be localized or delocalized in a photonic dimer \cite{28J Bloch coupled bistable} due to the interplay between the nonlinear interactions and interference when the laser energy is tuned across the antibonding/bonding modes. 

In the following, the second potential well 2 is excited using another laser beam with the same wavelength and spot size, during which the pumping flux of the potential well 1 is fixed at around 1.2 $P_{th}^1$. We monitor the emission of the polariton dimer by gradually increasing the pump density of the second laser beam from zero to above the threshold $P_{th}^2$ (here $P_{th}^2=$0.4 MW/cm$^2$ is the threshold of the condensation in the potential well 2). Surprisingly, the intensity of the polariton dimer decreases dramatically to nearly zero (Fig. \ref{fig:density} (a) and (c)) then increases again with the symmetric emission pattern appearing across the coupled potential wells, as shown in Fig. \ref{fig:density} (d). It is worth noting that the above results in Fig. \ref{fig:density} are reversible, i.e., when we gradually reduce the power of the second laser, the polariton dimer shows the same behavior at the same pump power.

\begin{figure}[t]
 \centering
 \includegraphics[width=\linewidth]{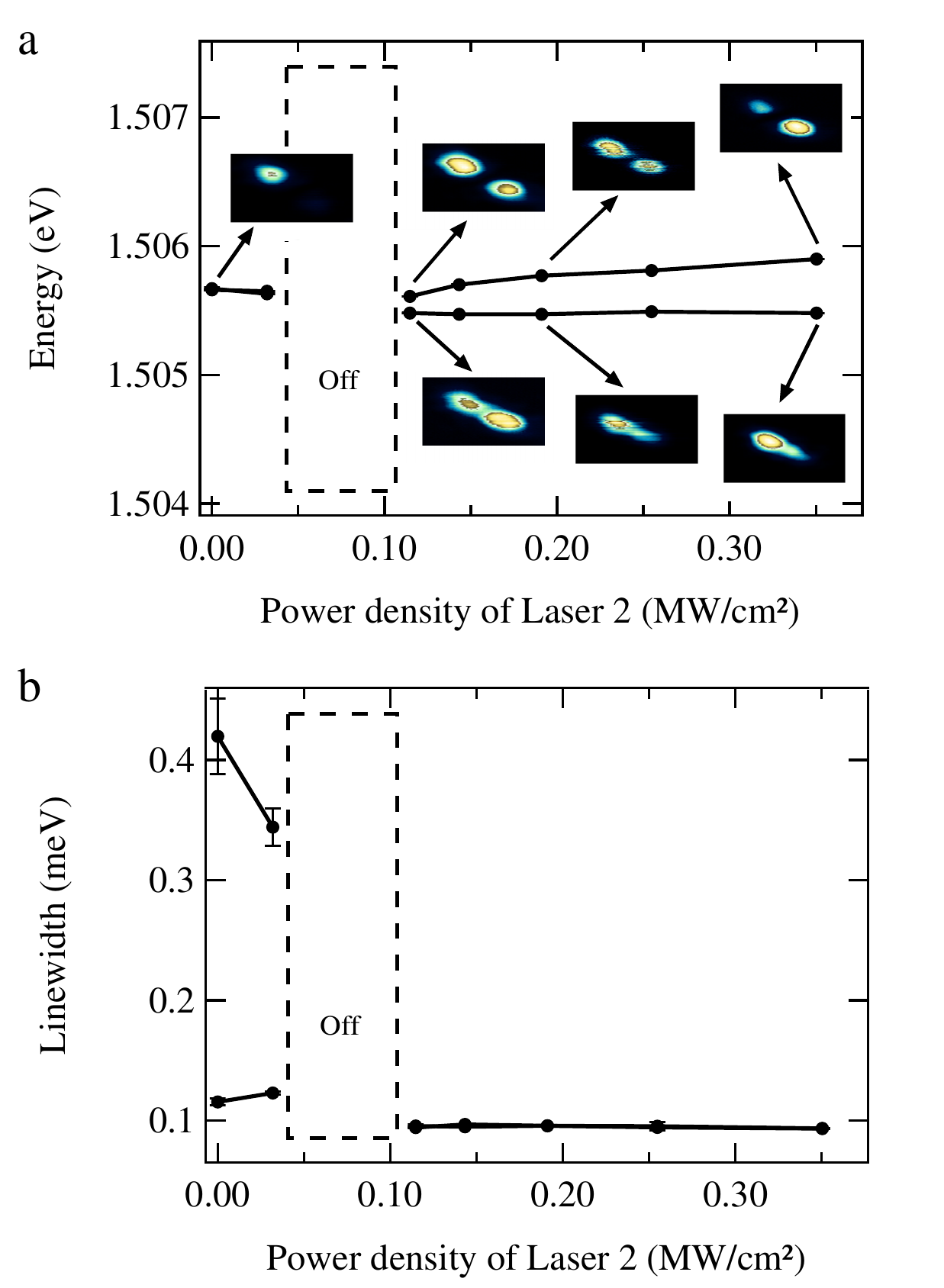} 
 \caption{\textbf{Demonstration of the exceptional point.} The real (a) and imaginary (b) part of the polariton energy of the coupled wells as the function of the pumping intensity of the potential well 2. The inserted graphs in (a) correspond to the real space imaging of the eigenmodes at several pumping intensities where the bottom ones are multiplied by 3 times, 0.2 times, 0.5 times respectively. The dashed regions show the switching off region. } \label{fig:EP}
\end{figure}

To check whether an exceptional point exists or not in the polariton dimer, we measure both the energy levels (real and imaginary parts) and the real space distribution of different polariton modes as the power of the second laser beam varies. Thanks to the coupling and the potential depth difference in the system, the polariton double-well potential can be tuned to the vicinity of the exceptional point by varying the pump power of the second laser beam. When the pump density of the potential well 2 is smaller than 0.032 MW/cm$^2$, there are two states whose energy share the same real part but different imaginary parts, as shown in Fig. \ref{fig:EP}. The total emission of the polariton dimer is mainly located in the potential well 1. At higher pumping density of the potential well 2, the polariton condensate is switched off and the coupled wells are below the threshold, which can be seen from the dispersion taken in the switching off regime in Fig. S3 of the SM, thus the emitted light intensity is greatly reduced. When the pump power density of the potential well 2 is larger than 0.115 MW/cm$^2$, the emission intensity of the polariton dimer increases sharply and we can find two states: the antibonding mode and bonding mode, as shown in the inserts in Fig. \ref{fig:EP} (a). The intensity of the mode with larger energy is much larger than another state thus dominates the total emission pattern (Fig. \ref{fig:density} (d)). The energy difference between these two states increases with the power of the second laser beam (Fig. \ref{fig:EP} (a)) and the intensity of the lower-energy state grows faster. At the same time, the linewidths of the two modes are around the same [Fig. \ref{fig:EP} (b)]. Hence, the clear bifurcations of the eigenenergies of the coupled condensates shown in Fig. \ref{fig:EP} are consistent with the theoretical prediction in Fig. \ref{fig:eigenenergy} (c,d), evidencing the existence of an exceptional point. In this scenario, the pump power of the second laser beam (the gain level difference between the two potential wells) acts as one parameter ($\Delta\gamma$) in the two dimensional parameter space in Fig. \ref{fig:eigenenergy}, which switches off the polariton condensate due to the substantial modification of the condensate wavefunction near the exceptional point. Under higher power of the second laser, the system approaches symmetric pumping, the emission of the polariton dimer becomes mainly localized at each potential site. In the coupled classical laser systems, we note that the appearance of two modes after switching off the coupled lasers or not also depends on the modal cross-saturation or spatial hole burning effect \cite{16reverse pump experiment}. 

If the pump power of the potential well 1 is increased, the polariton condensate can be partly switched off by repeating above experimental steps. For example, if the pump density of the potential well 1 is kept at around 2 $P_{th}^1$, the second laser can reduce the emission intensity at potential well 1 by about 40$\%$ at the pump density of 0.064 MW/cm$^2$ then enhance the intensity afterwards (see Fig. S5 in the SM). During this process, the emission intensity evolves from asymmetric to symmetric pattern. 

The coupled lasers can also be switched off when the frequencies of the two resonators is different \cite{16reverse pump experiment, 17reverse pump theory, 18lasing termination}. If we swap the two potential wells, that is, the potential well 2 is pumped above its threshold firstly and the pumping flux of potential well 1 is varied. In this configuration, the real parts of the polariton energy of the two potential wells are not the same due to the difference of the potential depths, however, the switching off polariton condensate can still be observed (see Fig.S6 in the SM). It means that in this case, the system experiences an anticrossing along a different route, instead of $\Delta{E}=0$, in Fig. \ref{fig:eigenenergy}(a,b).

%On another hand, the potential well A can be turned on directly if the pump density is around 0.8$P_{th}^1$ with increasing the pump density of the second laser beam. 
\section*{N\lowercase{umerical analysis}}
We numerically mimic the dynamics of the polariton condensate in our experiments by applying the extended Gross-Pitaevskii equation with loss and gain~\cite{29Wouters-PRL-2007}:
\begin{equation}\label{eq:GP}
\begin{aligned}
i\hbar\frac{\partial\Psi(\mathbf{r},t)}{\partial t}&=\left[-\frac{\hbar^2}{2m_{\text{eff}}}\nabla_\bot^2-i\hbar\frac{\gamma_\text{c}}{2}+g_\text{c}|\Psi(\mathbf{r},t)|^2 \right.\\
&+\left.\left(g_\text{r}+i\hbar\frac{R}{2}\right)n(\mathbf{r},t)+V(\mathbf{r})\right]\Psi(\mathbf{r},t), \\
\end{aligned}
\end{equation}
Here, $\Psi(\mathbf{r},t)$ is the polariton wavefunction, $m_{\text{eff}}$ denotes the effective mass of polaritons in the vicinity of the bottom of the lower polariton branch, $\gamma_\text{c}$ is the polariton loss rate, $g_\text{c}$ represents the polariton-polariton interaction, and $g_\text{r}$ represents the polariton-reservoir interaction. The reservoir $n$ is incoherent and provides the gain for the condensate with a rate $R$. It obeys the following equation of motion:
\begin{equation}\label{eq:reservoir}
\frac{\partial n(\mathbf{r},t)}{\partial t}=\left[-\gamma_r-R|\Psi(\mathbf{r},t)|^2\right]n(\mathbf{r},t)+P(\mathbf{r})\,.
\end{equation}
Here, $\gamma_\text{r}$ is the loss rate of the reservoir and $P(\mathbf{r})$ represents the two non-resonant pumps with Gaussian distribution, i.e., 
\begin{equation}\label{eq:pump}
P(x,y)=P_{1}e^{-\frac{(x-x_1)^2+y^2}{w^2}}+P_{2}e^{-\frac{(x-x_2)^2+y^2}{w^2}},
\end{equation}
with pump intensity $P_{1}$ and $P_{2}$ and pump width $w$. The two pumps are placed at $x_1$ and $x_2$, respectively. The double-well potential $V(\mathbf{r})$ takes the following form:
\begin{equation}\label{eq:potential}
V(x,y)=V_{\text{1}}e^{-\left(\frac{(x-x_1)^2+y^2}{W^2}\right)^2}+V_{\text{2}}e^{-\left(\frac{(x-x_2)^2+y^2}{W^2}\right)^2}.
\end{equation}
Here, $V_{1}$ and $V_{2}$ are the depths of the potential well 1 and 2, respectively, and $W$ indicates the size of each well. The parameters used for numerical modeling are summarized in \cite{30parameters}. We fix the intensity of pump 1 that excites potential well 1 with $P_1=12$ $\mu$m$^{-2}$~ps$^{-1}$ ($\sim1.2~P_{th}^1$) and gradually increase the other pump intensity $P_2$ in potential well 2, i.e., numerically repeating the experimental excitation process presented in Fig. \ref{fig:density}(a-d). The numerical results are shown in Fig.~\ref{fig:density}(e-g) (here the density profiles are integrated over time to match the experimental measurements) and are in very good agreement with the experimental results in Fig. \ref{fig:density}(b-d).

\section*{D\lowercase{iscussion}}

We note that the exciton polariton condensate in the polariton dimer can not be switched off under uniform pumping where the emission intensity continuously increases with the pumping flux, as shown in Fig. S7 of the SM. In this case the gain and loss coefficient of the two potential wells are modulated simultaneously. In addition, as we also find in our numerical simulations, the coupling between the two potential wells plays a critical role in the switching off process, which can not be observed if we use two tightly focused laser spots to excite a simple planar microcavity (see Fig. S8 in the SM). 

To summarize, the counter-intuitive results reported in the present work are rooted in the optically induced gain and loss modulation near an exceptional point in a polariton dimer. We demonstrate that this can be used to efficiently control the polariton condensation process and explore the role of the exceptional point in detail. Switching off the polariton condensate with increasing total pump power can also be used to investigate further the bistability or multistability \cite{31PT and multistability} in the non-Hermitian double-well potentials if the detuning between the excitons and cavity mode are tuned to be more positive. If the polariton condensate were loaded in multiple coupled potential wells, high-order exceptional points could be explored where the polariton energy is very sensitive against the perturbation\cite{11high order EP CDN,12high order EP yanglan}. For the future, we also envision that novel edge modes based on polariton condensates can be created in 1D or 2D potential lattices when the gain and loss coefficients are modulated similarly to the demonstration in the present work \cite{Lee edge non hermitian,topological invariants  wangzhong}.

\section*{A\lowercase{cknowledgments}}
%\begin{acknowledgments}
T.G.thanks Li Ge for fruitful discussion and acknowledges the support from the National Natural Science Foundation of China (grant No. $11874278$). The Paderborn group acknowledges the Deutsche Forschungsgemeinschaft (DFG) through the collaborative research center TRR142 (project A04, grant No. 231447078) and Heisenberg program (grant No. 270619725). X.M. further acknowledges support from the National Natural Science Foundation of China (grant No. 11804064). P.S thanks Project No. 041020100118 supported by Westlake University and Program 2018R01002 supported by Leading Innovative and Entrepreneur Team Introduction Program of Zhejiang. P.S. acknowledges financial support from Russian Science Foundation Grant No. 19-72-20120 for supporting MBE sample growth and from Greece Polisimulator project co-financed by Greece and EU Regional Development Fund.
%\end{acknowledgments}

\section*{M\lowercase{ethods}}

The setup we used in the experiment is a hand-made momentum-space spectroscopy system. The laser is divided into two identical beams through a 50:50 beam splitter, and is focused onto the sample through the same objective (X 50, NA 0.42). The intensity of each laser can be adjusted continuously. The signal emitted from the sample is collected by the objective and analyzed by a spectrometer (Princeton instrument), from which we can get the energy-momentum spectrum and energy-position spectrum of the exciton polaritons. The lens closest to the spectrometer is installed on a one-dimensional motorized stage (NewPort), and tomography can be performed (Fig. 4) by continuously moving the translation stage from which we can get the energy-resolved real space distribution of polaritons.

\textbf{Author contributions.} 
T.G. and X.M. conceived the project, Y.L. performed the experiment and analyzed the results, X.M. performed the theoretical analysis and numerical simulation, T.G. and X.M. prepared the manuscript with contributions from S.S. and P.S., Z. H and P.S. fabricated the sample. All authors discussed the results.

\textbf{Competing interests.}
The authors declare no competing interests.

%plain tex reference style


\begin{thebibliography}{99}
\bibitem{30parameters} The values of the parameters used for the numerical modeling are: $m_{\text{eff}}=5\times{10^{-5}~m_{\text{e}}}$ ($m_{\text{e}}$ is the free electron mass), $\gamma_\text{c}=0.16$~ps$^{-1}$, $\gamma_\text{r}=1.5~\gamma_\text{c}$, $g_\text{c}=6~\mu${eV}~$\mu${m}$^2$, $g_\text{r}=2g_\text{c}$, $R=0.01$~ps$^{-1}~\mu${m}$^2$, $w=1~\mu$m, $W=1.5~\mu$m, $V_{\text{1}}=-2.2$~meV, $V_{\text{2}}=-2$~meV, $x_1=-2~\mu${m}, and $x_2=2~\mu${m}.
\end{thebibliography}
\end{document}